# Tolerance against conducting filament formation in nanosheet-derived titania thin films

Masaya Sato,[1] Masahiro Hara,[2] Asami Funatsu,[2] and Ryo Nouchi[1,3]

[1] Department of Physics and Electronics, Osaka Prefecture University, Sakai 599-8570, Japan
[2] Faculty of Advanced Science and Technology, Kumamoto University, Kumamoto 860-8555, Japan
[3] PRESTO, Japan Science and Technology Agency, Kawaguchi 332-0012, Japan

E-mail: r-nouchi@pe.osakafu-u.ac.jp

**Abstract**

Herein, titania thin films are fabricated by a facile liquid-phase method based on vacuum filtration of a colloidal suspension of titania nanosheets, which is followed by thermal annealing to transform the nanosheet film into anatase $TiO_2$. Nanosheet-derived titania thin films exhibit poor resistive switching with an interface-type mechanism. This behaviour is distinct from the filamentary switching that has been observed with titania thin films fabricated by other conventional techniques. This tolerance against conducting-filament formation may be ascribed to a low concentration of oxygen vacancies in nanosheet-derived films, which is expected because of the O/Ti ratio of titania ($Ti_{0.87}O_2$) nanosheets being larger than that of $TiO_2$. Besides, the dielectric breakdown strength of nanosheet-derived films is found to be comparable to or higher than that of titania thin films fabricated by other techniques. These findings clearly indicate the usefulness of nanosheet-derived titania thin films for dielectric applications.

## 1. Introduction

Metal oxides are used in modern electronics as a high-κ dielectric for miniaturized transistors [1-3] and as a switching medium for resistance change memories [4-6] including memristors [7-9]. Among various metal oxides, titania [10-12] is significantly important as it is ubiquitous, which is due to the 10th highest Clarke number of Ti. Several methods have been reported for the fabrication of titania films, namely, physical vapour deposition (PVD)-based methods like sputtering [13,14], chemical vapour deposition (CVD)-based methods like atomic layer deposition (ALD) [15,16], and liquid-phase methods like sol-gel synthesis [17,18]. Among the reported methods, liquid-phase approaches require only simple fabrication methods without any sophisticated vacuum technology.

In this article, vacuum filtration of a colloidal suspension of titania nanosheets is employed as a facile liquid-phase method for the fabrication of titania thin films. Vacuum filtration methods have been widely applied for the thin film formation of various nanosheets ranging from chalcogenides [19] to oxides [20]. The fabricated thin film is found to exhibit a dielectric breakdown strength comparable to or higher than that of titania thin films fabricated by PVD and CVD based methods. In addition, the nanosheet-derived thin films with photolithographically patterned Au electrodes exhibit resistive switching, which can be explained using the framework of the interface-type mechanism. However, in Au-contacted titania thin films fabricated by other methods, the switching is interpreted using the filamentary mechanism. This tolerance of the nanosheet-derived titania thin films to the formation of conducting filaments may be attributed to the difference in the crystal structure between the titania nanosheet and conventional $TiO_2$. The O/Ti ratio of the nanosheet is larger than 2 [21,22]. Therefore, the number of oxygen vacancies should be suppressed in the nanosheet-derived titania films. The present finding indicates the suitability of nanosheet-derived titania films for dielectric applications.

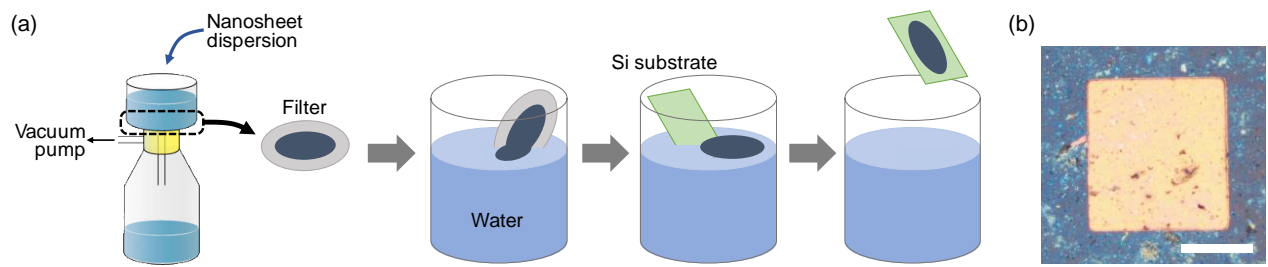

**Figure 1**. Fabrication of titania nanosheet thin films. (a) Procedure of vacuum filtration and subsequent transfer to a Si substrate. (b) Optical micrograph of a fabricated device with a top Au electrode. Scale bar: 100 μm.

## 2. Experimental

A colloidal suspension comprising of titania nanosheets was fabricated from a parent crystal ($K_{0.80}Ti_{1.73}Li_{0.27}O_4$) by following the procedure described in the literature [23]. Titania nanosheet thin films were fabricated by vacuum filtration and subsequent transfer to a target substrate [19]. The fabrication process of titania nanosheet thin films is illustrated in figure 1(a). First, the suspension was filtrated onto a filter membrane with nanoscale pores by vacuum filtration. Second, the membrane was immersed into ultrapure water, which left a nanosheet thin film floating at the water/air interface. Third, the floating film was scooped out using a $p^+$-Si substrate with a natural oxide layer with a thickness of at most 2 nm [24]. The $p^+$-Si substrate served as a bottom electrode for electrical characterizations.

The so-obtained titania nanosheet thin films on a $p^+$-Si substrate were annealed in air at 800°C for 5 h. A photolithographically patterned 30-nm-thick Au electrode was fabricated onto the annealed film by thermal evaporation and a subsequent lift-off process, as shown in figure 1(b). To strengthen the adhesion of Au to the film, an ultrathin Cr layer was pre-deposited onto the photolithographically defined resist pattern. The average thickness of the Cr adhesion layer was only 1 nm, which ensured the formation of an island-like form of the Cr layer. Thus, the interface between the top electrode and the titania film was dominated by Au.

Surface topographic images of the fabricated films were acquired by an atomic force microscope (AFM) in the dynamic force mode (Hitachi High-Technologies, AFM5200S). Raman scattering spectra of the films were acquired in ambient air by a Raman microscope (Nanophoton, Raman-DM) that was equipped with a 532-nm laser. The acquired Raman scattering spectra were calibrated using a Si-related peak at 520 cm$^{-1}$. The thickness of the films was determined by a surface profiler (Dektak150, Veeco). Electrical characterizations were conducted with a semiconductor device analyser (Keysight, B1500A) at room temperature in ambient conditions; the current compliance was set to 5 mA.

## 3. Results and discussion

Figure 2(a) shows an AFM image of a titania nanosheet thin film after annealing. This annealing process removed the organic impurities within the film (e.g., organic dispersant), and thus, AFM images were easily acquired without cantilever contamination by organic impurities on the surface. The surface roughness was found to be approximately 10 nm from the height profile; it is expected to be reduced by optimizing the filtration conditions. Another important effect of annealing is the transformation of the nanosheet film into the crystalline anatase phase of $TiO_2$ [25]. The Raman scattering spectrum in figure 2(b) exhibits a clear peak at 144 cm$^{-1}$, which is characteristic of anatase $TiO_2$ [26,27].

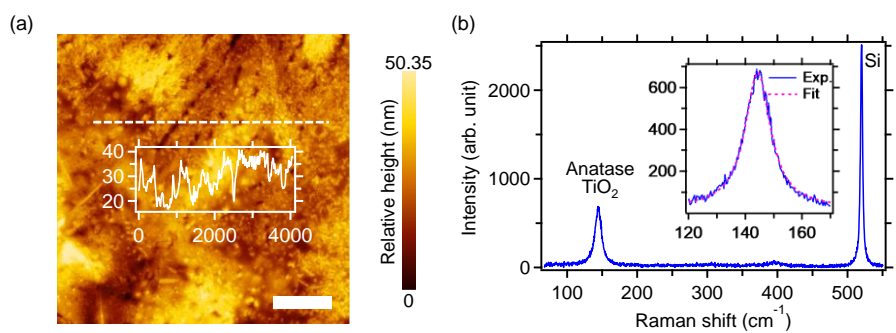

**Figure 2**. Characterization of the nanosheet-derived titania thin film. (a) AFM image of the transferred film after annealing. Scale bar: 1 μm. The inset shows a line profile along the white dashed line. (b) Raman scattering spectrum of the film after annealing. The inset shows the Lorentzian fit of the peak at 144 cm$^{-1}$.

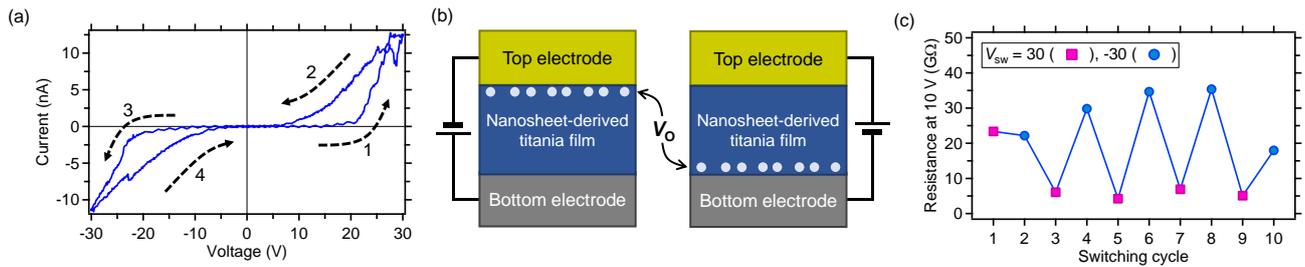

**Figure 3**. Resistive switching of the nanosheet-derived titania thin film with a thickness of 200 nm. The top and bottom electrodes were Au and p$^+$-Si, respectively. (a) Hysteretic current–voltage characteristics. (b) Interface-type mechanism accountable for the current loop in (a). $V_O$ denotes the oxygen vacancy. (c) Switching cycle obtained after application of the switching voltage, $V_{sw}$, for 1 s.

Figure 3(a) shows the current–voltage characteristics of the fabricated two-terminal Au/titania/p$^+$-Si device with a 200-nm-thick titania layer. The voltage was swept as indicated by arrows. During the sweep from 0 to +30 V (sweep 1), an increase in current was observed at 20 V, which corresponds to switching to a low-resistance state (LRS). The mechanisms of resistive switching can be classified into two types: namely, filament and interface types [4-6]. As evidenced from the sweep from +30 to −30 V (sweeps 2 and 3), the device clearly shows a rectified current–voltage curve and subsequently exhibits switching to the LRS at around −20 V, which is characteristic of the interface-type resistive switching [5,28].

Oxygen vacancies are known to play a dominant role in both the switching mechanisms. In the filamentary mechanism, switching to the LRS occurs when oxygen-deficient low-resistance filaments form bridges between the top and bottom electrodes. In the interface-type mechanism, positively charged oxygen vacancies move toward the interface with the negatively biased electrode, as schematically shown in figure 3(b). The positive charges, together with their image charges in the electrode, form an electric double layer at the electrode surface; this leads to the lowering of the effective work function of the electrode. As a result, electron injection from the electrode is enhanced and the device switches to the LRS.

The nanosheet-derived titania thin films with the Au contact were found to show interface-type resistive switching. As shown in figure 3(a), the switching was almost symmetric, indicating that the switching occurred at both electrode interfaces in a similar way. In addition, these films show poor switching as shown in figure 3(c), where the incomplete switching of the initial plot may be attributed to the incomplete movement of oxygen vacancies due to the short duration of the switching voltage application. On the contrary, Au-contacted titania thin films formed by other methods have been reported to show filamentary resistive switching with high switching ratios [29-34]. A plausible explanation of this difference is a lower expected density of oxygen vacancies in nanosheet-derived films. The starting material of the present titania films is a titania nanosheet having chemical formula $Ti_{0.87}O_2$ [22], and therefore, the O/Ti ratio is larger than 2. Thus, the two-dimensional nanosheets are more oxygen-rich than three-dimensional $TiO_2$. Consequently, the concentration of oxygen vacancies in the nanosheet-derived titania films is expected to be small because of the oxygen-rich condition during the transformation of nanosheets to anatase. Conducting filament formation is known to be efficient in defective titania films [33] and devices with oxygen-absorbing metals [30]. Therefore, the tolerance of the nanosheet-derived titania thin films against conducting filament formation may be due to a low concentration of oxygen vacancies.

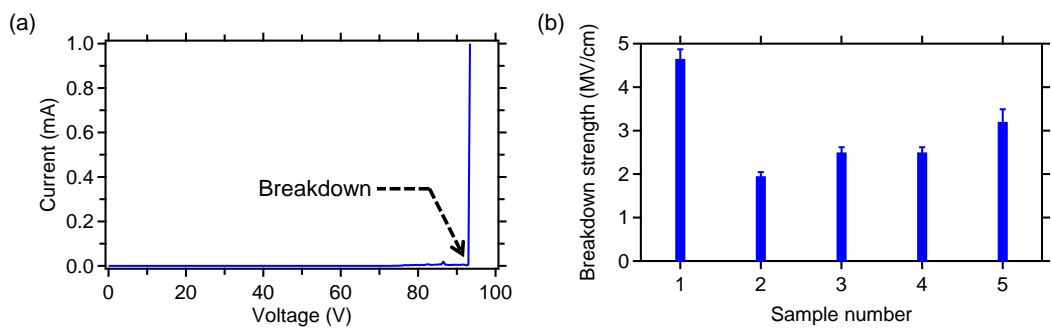

**Figure 4**. Dielectric breakdown of the nanosheet-derived titania thin film. (a) Current–voltage characteristics of the device in figure 3. (b) Breakdown strength measured with five devices. The error bars originate from the thickness inhomogeneity. The thicknesses of the titania films were ~200 nm for No. 1–4 and ~100 nm for No. 5.

It is known that the higher the crystallinity of the sample, the sharper the Raman peak. Therefore, the full width at half maximum (FWHM) of the anatase peak near 144 cm$^{-1}$ was analysed in detail. The inset in figure 2(b) shows the fitting curve using the Lorenz function. The fitting curve shows a FWHM of 11 cm$^{-1}$. This is significantly narrower than the FWHM of titania films that show filament switching (17 cm$^{-1}$ and 19 cm$^{-1}$ for Ref. [35] and Ref. [36], respectively). This indicates that our anatase titania film derived from nanosheets actually has a higher crystallinity.

Figure 4 shows the test of dielectric breakdown of the nanosheet-derived films. An electric field required to induce the breakdown was determined from the current–voltage curve as shown by an arrow in figure 4(a). The determined breakdown strength ranges from 2 to 5 MV/cm as shown in figure 4(b). This range is comparable to or slightly higher than the breakdown strengths of sol-gel-synthesized films (1—5 MV/cm) [17], ALD-grown films (0.5 MV/cm) [15], and sputter-grown films (0.5—2 MV/cm) [13]. This breakdown behaviour was irreversible and the state after the breakdown could not be recovered by applying a high voltage with the opposite polarity. Thus, this dielectric breakdown is distinct from the conducting filament formation that is reversible in nature. These findings again suggest the tolerance of the nanosheet-derived titania films against conducting filament formation.

## 4. Conclusions

In this study, we investigated the resistive switching behaviour and dielectric strength of titania thin films, which were fabricated by a facile liquid-phase method based on the vacuum filtration of a colloidal suspension of titania nanosheets. The nanosheet-derived thin films after annealing were confirmed to correspond to the anatase phase of $TiO_2$. They exhibited a poor resistive-switching behaviour with the interface-type mechanism, which was distinct from the filamentary mechanism observed in titania thin films fabricated by other conventional techniques. This is due to the O/Ti ratio of titania ($Ti_{0.87}O_2$) nanosheets being larger than the stoichiometric ratio of $TiO_2$. The O/Ti ratio being larger than 2 should have resulted in a low concentration of oxygen vacancies after the transformation into anatase. Since oxygen vacancies are necessary for resistive switching, the low concentration suppressed the switching in the nanosheet-derived films. The dielectric breakdown strength of the nanosheet-derived films was confirmed to be comparable to or slightly higher than those of the titania thin films fabricated by other conventional techniques. These findings indicate the superiority of the nanosheet-derived titania thin films for dielectric applications.

## Acknowledgements

This work was supported by JSPS KAKENHI Grant Numbers JP17H01040 and JP19H02561; and JST, PRESTO Grant Number JPMJPR17S6.

## Data availability

The data that support the findings of this study are available from the corresponding author upon reasonable request.